\pdfoutput=1
\RequirePackage[l2tabu,orthodox]{nag}
\documentclass[a4paper,11pt]{article}
\usepackage[T1]{fontenc}
\usepackage[utf8]{inputenc}
\usepackage{pos}
\usepackage{microtype}
\usepackage{mathtools}
\usepackage{csquotes}
\usepackage{physics}
\usepackage{adjustbox}
\usepackage{slashed}
\usepackage{tikz}
\usepackage[outline]{contour}
\contourlength{2.5pt}
\usepackage{simpler-wick}
\usepackage{booktabs}
\usepackage{siunitx}

\usepackage[style=base]{caption}
\usepackage{subcaption}
\NewDocumentCommand\cpi{}{\uppi}                        % pi constant (upright symbol)
\NewDocumentCommand\I{}{\mathrm{i}}
\NewDocumentCommand\e{ m }{\mathrm{e}^{#1}}
\NewDocumentCommand\Op{}{\mathcal{O}}

\usepackage{bm}
\RenewDocumentCommand\vec{ m }{\bm{\mathrm{#1}}}

\NewDocumentCommand\Nc{}{N_{\mathrm{c}}}
\NewDocumentCommand\Nf{}{N_{\mathrm{f}}}
\NewDocumentCommand\gE{}{g_{\mathrm{E}}}
\NewDocumentCommand\mE{}{m_{\mathrm{E}}}
\DeclareDocumentCommand\textoverline{ m }{$\mathsurround=0pt\cramped{\overline{\hbox{#1}}}$}
\NewDocumentCommand\MSbar{}{\textoverline{\textsc{ms}}}

\DeclareSIUnit\fm{\femto\metre}

\title{Hadronic screening masses in thermal QCD up to the electroweak scale}
\author*[a,b]{Marco Cè}
\author[a,b]{Leonardo Giusti}
\author[b]{Michele Pepe}
\author[a,b]{Pietro Rescigno}

\affiliation[a]{Dipartimento di Fisica \enquote{G. Occhialini}, Università degli Studi di Milano-Bicocca}
\affiliation[b]{Istituto Nazionale di Fisica Nucleare (INFN), Sezione di Milano-Bicocca}
\emailAdd{marco.ce@unimib.it}

\abstract{%
  Novel theoretical and computational strategies have opened the possibility of exploring thermal QCD at the non-perturbative level at unprecedented temperatures, reaching from the \si{\GeV} scale up to the electroweak scale.
  A number of observable quantities are now being investigated in this regime.
  Key ones are the hadronic screening masses, which encode the correlation length of the medium and thus the extent to which strong interactions are screened in a thermal environment.
  In these proceedings we present recent lattice results for hadronic screening masses, including baryonic modes and preliminary non-static mesonic modes.
  These results can be compared with predictions from the perturbative expansion in the three-dimensional effective theory valid at asymptotically large temperatures.
  The comparison reveals persistent higher-order effects including those of non-perturbative origin, up to the electroweak scale, shedding new light on the microscopic structure of QCD at extreme temperatures.
}

\FullConference{The 42nd International Symposium on Lattice Field Theory (LATTICE2025)\\
2-8 November 2025\\
Tata Institute of Fundamental Research, Mumbai, India\\}

\begin{document}
\maketitle

\section{Introduction}

QCD at finite temperature is crucial for many areas of physics, including relativistic heavy ion collision experiments and early universe cosmology \cite{Laine:2016hma,Giusti:2024ohu,Ce:2025ihd}.
While perturbation theory is available at high temperatures, Linde's problem implies that perturbative expansions in thermal QCD are computable only up to a finite order in the QCD coupling.
Consequently, a non-perturbative approach is required for a full understanding of the system \cite{Linde:1980ts,Ginsparg:1980ef,Appelquist:1981vg}.

In these proceedings, we outline two specific aspects of mesonic screening masses at very high temperatures $T$, namely the \emph{hyperfine} splitting~\cite{Ce:2025ihd} and screening masses in the non-zero Matsubara sectors~\cite{Ce:2026inprep}.
We study these both using effective field theory methods and lattice simulations of $\Nf=3$ QCD.

\subsection{Effective field theory description of QCD at high \texorpdfstring{$T$}{temperature}}

At asymptotically high temperatures, QCD exhibits a hierarchy of three thermal scales: the hard scale ($\cpi T$), the soft scale ($\mE\propto gT$), and the ultra-soft scale ($\gE^2\propto g^2T$)~\cite{Linde:1980ts,Braaten:1995jr,Kajantie:1995dw,Kajantie:1997pd}, see ref.~\cite{Laine:2016hma} for a review.
Indeed, thanks to the asymptotic freedom of the theory, $T\gg\Lambda_{\textrm{QCD}}$ implies $g(T)/\cpi \ll 1$ and in turn
\begin{equation}
  \frac{\gE^2}{\cpi} \ll \mE \ll \cpi T .
\end{equation}
Integrating out high-energy modes leads to a dimensionally-reduced effective field theory description, which assumes different names based on the degrees of freedom left in the theory.
Integrating out all fermionic and bosonic modes, which have energies of at least $\cpi T$, and all bosonic modes except for the lowest one, we are left with chromomagnetic $A_i$ and chromoelectric $A_0$ gluonic $n=0$ modes and no dependence on the imaginary time coordinate.
This is electrostatic QCD (EQCD), a three-dimensional non-Abelian $\mathrm{SU}(3)$ gauge theory with a dimensionful gauge coupling $\gE$ and a mass $\mE$ acquired by the adjoint scalar field $A_0$.
Tree-level matching to QCD results in~\cite{Kajantie:1997tt,Laine:2005ai}
\begin{equation}
  \gE^2 = g^2 T + \order{g^4 T^2}, \qquad \mE^2 = g^2 T^2 \left( \frac{\Nc}{3} + \frac{\Nf}{6} \right) + \order{g^4 T^2} .
\end{equation}
At this level in the matching, for $g/\cpi \ll 1$ the mass $\mE$ is parametrically larger than the $\order*{\gE^2}$ ultra-soft physics, and the $A_0$ field can be further integrated out to study distance scales longer than $1/\mE$.
This leaves us with a theory of only the chromomagnetic fields $A_i$, referred to as magnetostatic QCD (MQCD).
MQCD is actually a three-dimensional Yang-Mills theory, and quantities in MQCD have to be computed with non-perturbative methods such as 3d lattice simulations~\cite{Linde:1980ts,Appelquist:1981vg,Kajantie:1997tt}.

\subsection{Lattice methodology}

Although QCD at any temperature can be studied with standard lattice methods and four-dimensional simulations, the usual approach of setting the scale with a hadronic scheme may not be feasible.
In fact, at finite $T$ one has to satisfy
\begin{equation}
  L^{-1} \ll M_{\textrm{hadron}} \ll T < a^{-1},
\end{equation}
leading to a \emph{window problem} if the temperature is much higher than the hadronic scale $M_{\textrm{hadron}}$, since the system size becomes exceedingly large in lattice units $L/a$.
To overcome this, we set the scale using the non-perturbatively renormalized Schrödinger-functional (SF) or gradient-flow (GF) coupling at $\mu \approx T$ \cite{Luscher:1991wu,Bruno:2017gxd}.
This allowed for simulations of massless $\Nf=3$ QCD spanning from $T\approx\SI{1}{\GeV}$ up to \SI{160}{\GeV}, see ref.~\cite{DallaBrida:2021ddx} for all details.

We also employ shifted boundary conditions \cite{Giusti:2011kt, Giusti:2012yj}, with shift vector always chosen as $\vec{\xi}=(1,0,0)$.
While shifted boundary conditions are crucial for the calculation of the equation of state studied in refs.~\cite{Bresciani:2025vxw,Bresciani:2025mcu}, they are not necessary for computing screening masses.
Here, we employ them because those other studies and this project use the very same gauge field configurations in order to share the computational cost of generating them.
As a by-product, in the study of screening masses we observe reduced cut-off effects~\cite{DallaBrida:2021ddx}.

For the $n$th Matsubara sector, the screening masses are the exponential rate of decay of the two-point correlator $C^{(n)}_{\Op}(x_3)$ at asymptotic spatial distances
\begin{equation}
  m^{(n)}_{\Op} = -\lim_{x_3\to\infty} \dv{}{x_3} C^{(n)}_{\Op}(x_3)
\end{equation}
\begin{equation}
  C^{(n)}_{\Op}(x_3-y_3) = \int_{0}^{1/T} \dd{x_0} \int \dd{x_1} \dd{x_2} \, \e{\I\omega_n x_0} \left\langle \Op(x) \Op^\dagger(y) \right\rangle \simeq c^{(n)}_{\Op} \exp{-m^{(n)}_{\Op} \abs{x_3-y_3}} ,
\end{equation}
where the Matsubara frequency $\omega_n$ is $2n\cpi T$ for bosons and $(2n+1)\cpi T$ for fermions.

At variance with previous work where we employed point sources, in this computation we use stochastic wall sources with $\mathrm{U}(1)$ noise and fixed source $y_3$ coordinate.
We found that wall sources are very effective at reducing the statistical error on the correlator through volume averaging, without the noisy estimator significantly contributing to the error.
This is a consequence of the $L_0\times (288a)^2$ volume of our wall sources, very large in units of the correlation lengths $\propto 1/T=L_0\in\{4,6,8,10\}a$. 
Interestingly, at high temperatures we observe equally reduced errors for all the mesonic channels including the vector one.
This is significant because at $T=0$ the vector channel with wall sources empirically shows a much higher level of noise than, for instance, the pseudoscalar one.
Instead, at temperatures much higher than the chiral crossover the physics of screening correlators is dominated by the same $\sim\cpi T$ modes, and in particular the four-point contribution to the variance of stochastic estimator of two-point functions is very similar.
In the results presented here, we employed four wall sources with different noise and different source $y_3$ for each gauge field configuration.

\section{The hyperfine splitting}
\label{sec:hyperfine}

At high $T$, the pseudoscalar ($P=\bar{u}\gamma_5d$) and transverse vector ($V_T=\bar{u}\gamma_2d$) channels at $n=0$ have the lowest screening masses.
At asymptotically high temperatures these masses are calculable in the dimensionally-reduced theory, with quarks described by heavy fields.
This is referred to as three-dimensional non-relativistic QCD (3d NRQCD)~\cite{Laine:2003bd}.
Computing the screening masses involves obtaining the mesonic correlators, performing a perturbative expansion in the coupling of the effective theory $\gE^2=g^2 T + \dots$, and solving a $(2+1)d$ Schrödinger equation with the appropriate potential.
At $\order*{g^2}$ the potential is spin-independent and does not distinguish those two channels,
\begin{equation}\label{eq:leading_potential}
  U_0(\vec{r}) = \gE^2 \frac{C_{\mathrm{F}}}{2\cpi} \left[ \ln\frac{\mE r}{2} + \gamma_{\mathrm{E}} - K_0(\mE r) \right] , \quad r = \abs{\vec{r}}
\end{equation}
where $C_{\mathrm{F}}=(\Nc^2-1)/(2\Nc)$.
Thus, at this order in perturbation theory, $P$ and $V_T$ screening masses are degenerate~\cite{Laine:2003bd,Hansson:1991kb},
\begin{equation}\label{eq:static_perturbative}
  m^{(0)}_{V_T} = m^{(0)}_{P} = 2\cpi T \left( 1 + \num{0.032739961}\, g^2 \right) + \order{g^3} .
\end{equation} 
The leading spin-dependent contribution to the potential is temperature suppressed and the hyperfine splitting $\Delta m_{VP} = m^{(0)}_{V_T} - m^{(0)}_{P}$ starts at $\order*{g^4}$~\cite{Bala:2025ilf,Ce:2025ihd}
We report here the results of our calculation in ref.~\cite{Ce:2025ihd}, see also ref.~\cite{Bala:2025ilf} for an independent calculation of the same contribution.
The mesonic correlators in 3d NRQCD are expanded to next-to-leading order, keeping track of the temperature-suppressed contributions.
Among these, we are interested in the spin-dependent contribution to the potential since it contributes to the spin splitting,
\begin{equation}
  U_{\pm}(\vec{r}) = \mp \gE^2 \frac{C_{\mathrm{F}}}{(2\cpi T)^2} \delta^{(2)}(\vec{r}) , \quad \text{with $\pm$ for the $P$ and $V_T$ channel respectively} .
\end{equation}
We treat this as a perturbation of the $\order*{g^2}$ spin-independent potential $U_0$ in eq.~\eqref{eq:leading_potential}.
Since both potentials are symmetric under 2d rotations, we can solve the radial Schrödinger equation.
The perturbed energies are
\begin{equation}
  E_{\pm} = E_0 + \int_{\vec{r}} U_{\pm}(\vec{r}) \abs{\psi_0(\vec{r})}^2 \dd[2]{r} = E_0 \mp \frac{C_{\mathrm{F}}}{(2\cpi T)^2}  \abs{\psi_0(\vec{0})}^2 = E_0 \mp \frac{\gE^2\mE^2}{2} \frac{C_{\mathrm{F}}}{(2\cpi T)^2} \abs{\hat{\psi}_0(0)}^2 ,
\end{equation}
where $\hat{\psi}_0(\hat{r})=\psi_0(\vec{r})\sqrt{2\cpi}/\mE$, with $\hat{r}=\mE\abs*{\vec{r}}$, is the rescaled radial wave function of the lowest eigenmode of the Schrödinger equation.
Applying the leading-order matching to QCD, the spin-splitting is
\begin{equation}\label{eq:splitting}
    \frac{m_{VP}}{2\cpi T} = \frac{\gE^2\mE^2}{\cpi} \frac{C_{\mathrm{F}}}{(2\cpi T)^3} \abs{\hat{\psi}_0(0)}^2 = \frac{g^4}{4\cpi^4} \abs{\hat{\psi}_0(0)}^2 + \order{g^5} = \num{0.002376}\, g^4  + \order{g^5} .
\end{equation}

\begin{figure}[tb]
  \includegraphics[width=.32\columnwidth]{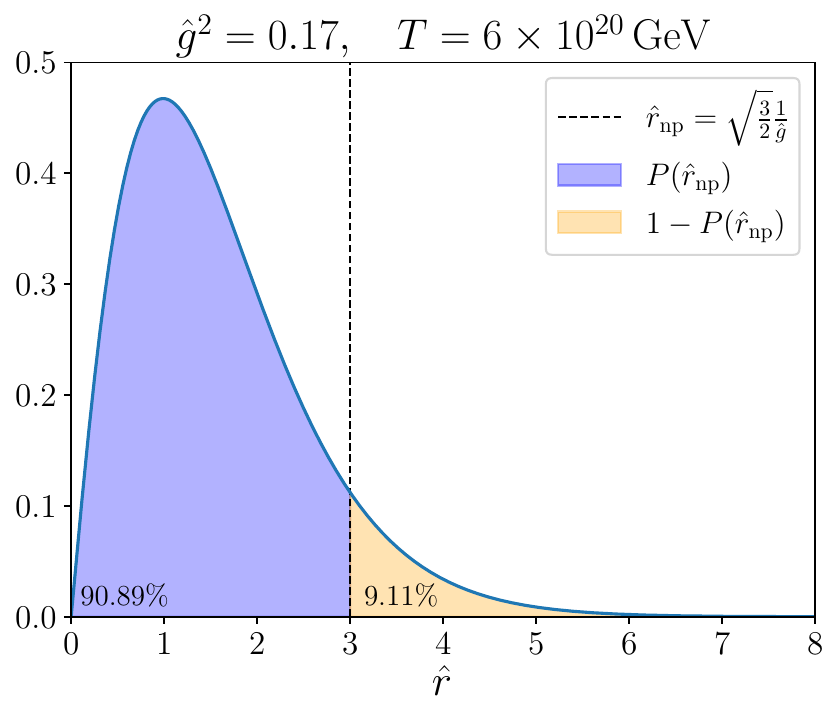}%
  \includegraphics[width=.32\columnwidth]{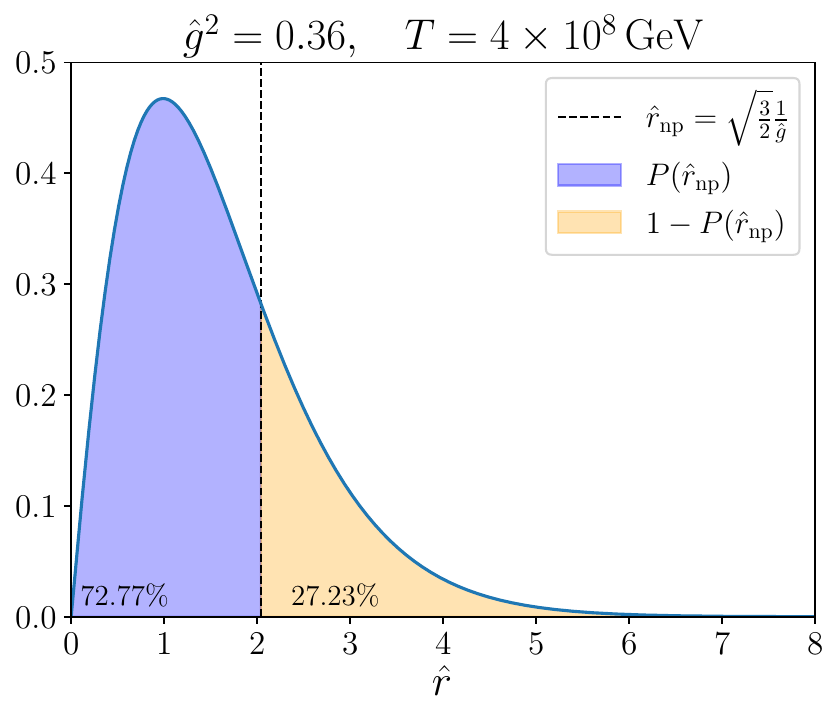}%
  \includegraphics[width=.32\columnwidth]{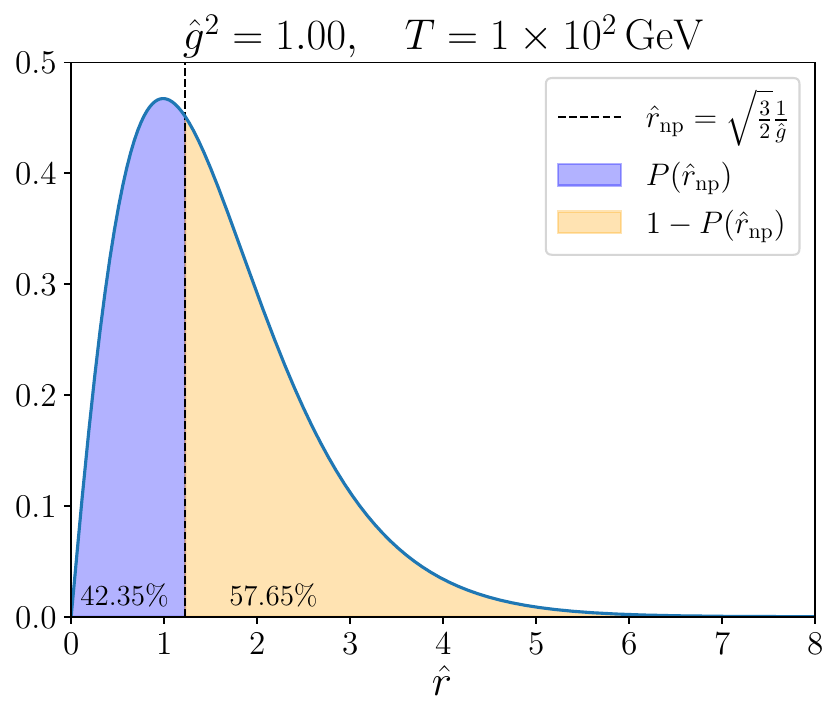}
  \caption{%
    Radial probability density of wave function.
    The vertical line indicates the radial coordinate at which non-perturbative contributions start to be relevant, $r_{\textrm{np}}=1/\gE^2\simeq 1/\sigma$, for three different values of the QCD coupling $\hat{g}^2=0.17$, $0.36$ and $1.00$, and corresponding temperatures $T=\SI{6e20}{\GeV}$, \SI{4e8}{\GeV} and \SI{100}{\GeV}.
  }\label{fig:wave_function_probability}
\end{figure}

This result requires a few considerations~\cite{Ce:2025ihd}.
The full potential is expected to include a string contribution which is dominant for large distances and non-perturbative in nature.
This contribution appears at $\order*{g^3}$ which in turn is $\order*{g^5}$ in the splitting of eq.~\eqref{eq:splitting}.
Thus, while the first spin-splitting contribution is calculable in perturbation theory, a non-perturbative approach is required to go beyond this result.
However, although the splitting results from an ultralocal interaction, a non-vanishing value of the wave function at the origin, eq.~\eqref{eq:splitting} introduces a dependence on the normalization of the wave function over all radial distances including large ones at which the string tensions dominates the full potential.
Denoting with $r_{\textrm{np}}=1/\gE^2\simeq 1/\sigma$ the threshold to these distances, it is instructive to assess to what extent the wave function probes radial distances at which the full non-perturbative potential is expected to be very different from the perturbative potential considered here.
The probability $P(\hat{r}>\hat{r}_{\textrm{np}})$ is of course a function of the temperature, and it is represented graphically in figure~\ref{fig:wave_function_probability}.
We observe that $P(\hat{r}>\hat{r}_{\textrm{np}})>\SI{50}{\percent}$ for any $T<\SI{100}{\GeV}$, and the probability falls below \SI{10}{\percent} only for asymptotically large temperatures larger than \SI{e20}{\GeV}.
This leads us to conclude that the perturbative $g^4$ contribution computed here is likely to account for most of the spin-splitting only at asymptotically-high temperatures, while ultra-soft non-perturbative contributions originating from the string tension and starting at $\order*{g^5}$ are likely to be relevant even up to the electroweak scale.

\begin{figure}[t]
  \centering%
  \includegraphics[width=.6\columnwidth]{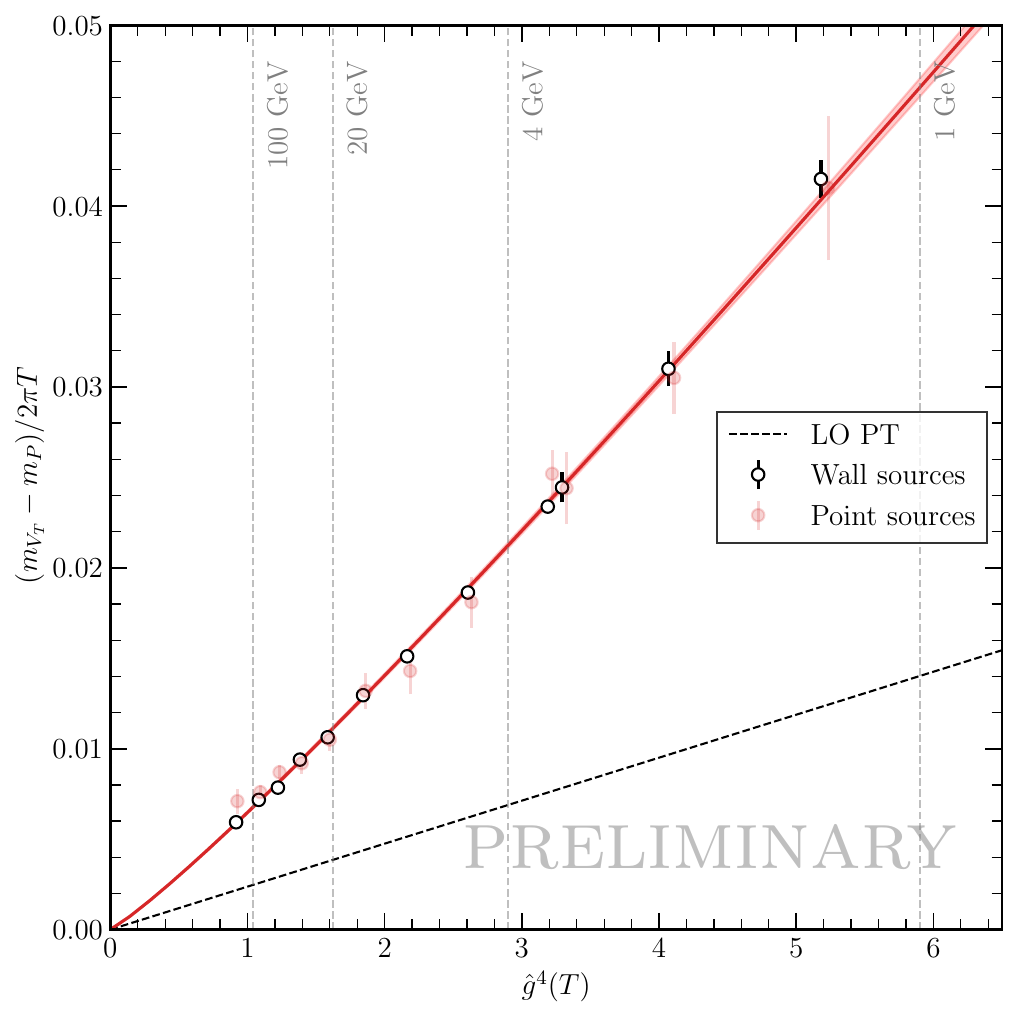}
  \caption{%
    Hyperfine splitting in the $n=0$ sector as a function of $\hat{g}^4$, with point-sources data published in ref.~\cite{DallaBrida:2021ddx} and preliminary wall-sources data presented here, compared to the leading-order perturbative spin-splitting result published in ref.~\cite{Ce:2025ihd} and discussed in section~\ref{sec:hyperfine}.
  }\label{fig:spin_splitting}
\end{figure}

As discussed in ref.~\cite{Ce:2025ihd}, this is supported by non-perturbative results for the $n=0$ screening masses published in ref.~\cite{DallaBrida:2021ddx}.
As a byproduct of the $n>0$ computation discussed in the next section, we have now available more precise results for the $P$ and $V_T$ screening masses at $n=0$, and thus for the spin splitting, that we can compare to the perturbative prediction.
These preliminary new results are shown in figure~\ref{fig:spin_splitting}, together with a preliminary fit as a function of $\hat{g}$ and the old data in a lighter shade.
In this plot, $\hat{g}$ is the two-loop \MSbar\ coupling at $\mu=2\cpi T$
\begin{equation}
  \frac{1}{\hat{g}^2(T)} \equiv \frac{9}{8\cpi^2} \ln{\frac{2\cpi T}{\Lambda_{\text{\MSbar}}}} + \frac{4}{9\cpi^2} \ln(2\ln{\frac{2\cpi T}{\Lambda_{\text{\MSbar}}}}) ,
\end{equation}
with $\Lambda_{\text{\MSbar}}=\SI{341}{\MeV}$ from ref.~\cite{Bruno:2017gxd}.
We choose $\hat{g}(T)$ because it is a convenient proxy for the temperature, suggested by the effective theory analysis, with the crucial point being the leading logarithmic dependence on $T$.
A few corresponding temperature values are indicated in the plots by dashed vertical grey lines.

With wall-sources data, the error on the continuum-extrapolated results at each temperature is reduced by up to one order of magnitude with respect the data from ref.~\cite{DallaBrida:2021ddx}.
At first glance, the spin splitting in figure~\ref{fig:spin_splitting} appears to grow roughly linearly with $\hat{g}^4$, but with an effective coefficient that is about three times the leading-order result in the dimensionally-reduced theory.
However, the updated data points are not compatible with a simple linear fit in $\hat{g}^4$, and higher order terms are required to describe the data.
A preliminary fit with $\Delta m_{VP}/2\cpi T=\num{0.002376}\,\hat{g}^4 + s_5\hat{g}^5 + s_6\hat{g}^6$ correctly describe the data, with preliminary values for the fit parameters $s_5=\num{0.00468(28)}$ and $s_6=\num{-0.00065(23)}$ and a strong autocorrelation between the two.
This parametrization can be used to obtain the hyperfine splitting at any temperature above \SI{1}{\GeV}.
At all the temperatures simulated on the lattice up to the highest of about \SI{160}{\GeV}, the sum of the higher-order contributions $s_5\hat{g}^5$ and $s_6\hat{g}^6$ is effectively larger than the leading-order $\order*{g^4}$ contribution computed in refs.~\cite{Bala:2025ilf,Ce:2025ihd}.
This supports the picture in which the leading $g^4$ behaviour is recovered only at asymptotically high temperature, while higher-order contributions (which include those of non-perturbative origin that starts at $\order*{g^5}$) are dominant even at the electroweak scale. 

\section{Mesonic screening masses for \texorpdfstring{$n>0$}{n>0}}
\label{sec:non-static}

Thanks to the reduced statistical errors resulting from the stochastic estimator with volume sources and $U(1)$ noise, we are able to investigate also the non-static Matsubara sector, $n>0$.
Here we focus on mesonic screening masses with $n=1$.
At leading order in 3d NRQCD we expect~\cite{Brandt:2014uda}
\begin{subequations}\label{eq:nonstatic_perturbative}
  \begin{gather}
    m_{V_0}^{(1)}|_{\textrm{PT}}                            = 2\cpi T \left( 1 + \num{0.0533477}\, g^2 \right) + \order{g^3} , \\
    m_{V_0}^{(0)}|_{\textrm{PT}}                            = 2\cpi T \left( 1 + \num{0.0718778}\, g^2 \right) + \order{g^3} , \\
    m_P^{(1)}|_{\textrm{PT}} = m_{V_T}^{(1)}|_{\textrm{PT}} = 2\cpi T \left( 1 + \num{0.0748404}\, g^2 \right) + \order{g^3} ,
  \end{gather}
\end{subequations}
where $V_0=\bar{u}\gamma_0d$ is the vector current in the temporal direction.
We observe that in the free theory both $n=0$ and $n=1$ screening masses are proportional to $2\cpi T$, but at leading order in $g^2$ pseudoscalar ($P$) and transverse vector ($V_T$) $n=1$ masses are heavier than their $n=0$ counterpart, while still being degenerate.
The screening masses in the $V_0$ channel are intermediate and they have the opposite ordering, with the $n=1$ mass being lighter than $n=0$.

In the right panel of figure~\ref{fig:summary} we compare as a function of $\hat{g}^2(T)$ these leading-order predictions with the continuum-extrapolated results of our high-$T$ lattice simulations for $m_{V_0}^{(1)}$, $m_{V_0}^{(0)}$ and $m_P^{(1)}$.
On the one hand these data points confirm the hierarchy of the $g^2$ contributions in eqs.~\eqref{eq:static_perturbative} and~\eqref{eq:nonstatic_perturbative}.
On the other hand all masses are consistently larger than the one-loop perturbative prediction, indicating non-negligible higher order contributions, which likely include non-perturbative effects.
As shown in the left panel of the same figure, the higher-order contributions have opposite sign with respect to the static $m_P^{(0)}$ and $m_{V_T}^{(0)}$ case.

The data for the $V_0$ channel confirms at the non-perturbative level the opposite ordering from perturbation theory, with the $n=1$ mass being lighter than $n=0$.

Data points for $m_{V_T}^{(1)}$ are not shown but they would overlap those for $m_P^{(1)}$: the difference $m_{V_T}^{(1)}-m_P^{(1)}$ is statistically compatible with zero.
A preliminary analysis in the effective theory indicates that the spin splitting in the $n=1$ sector vanishes up to and including $\order*{g^4}$.
Our computation confirms that also higher-order contributions are small compared to the $n=0$ hyperfine splitting of section~\ref{sec:hyperfine}.
Finally, we observe that $m^{(1)}_{V_0} = m^{(1)}_{V_3}$, with $V_3=\bar{u}\gamma_3d$, as expected from Ward identities.\footnote{%
  While $\partial_3 C^{(0)}_{V_3}(x_3)=0$ thus $m^{(0)}_{V_3}$ is not defined.
}
While not shown here, we also measured masses in the scalar ($S$) and axial vector ($A$) channels, and found them to be compatible with pseudoscalar and vector masses respectively, consistently with restoration of chiral symmetry at high $T$.
\section{Conclusions and outlook}

\begin{figure}[t]
  \includegraphics[width=\columnwidth]{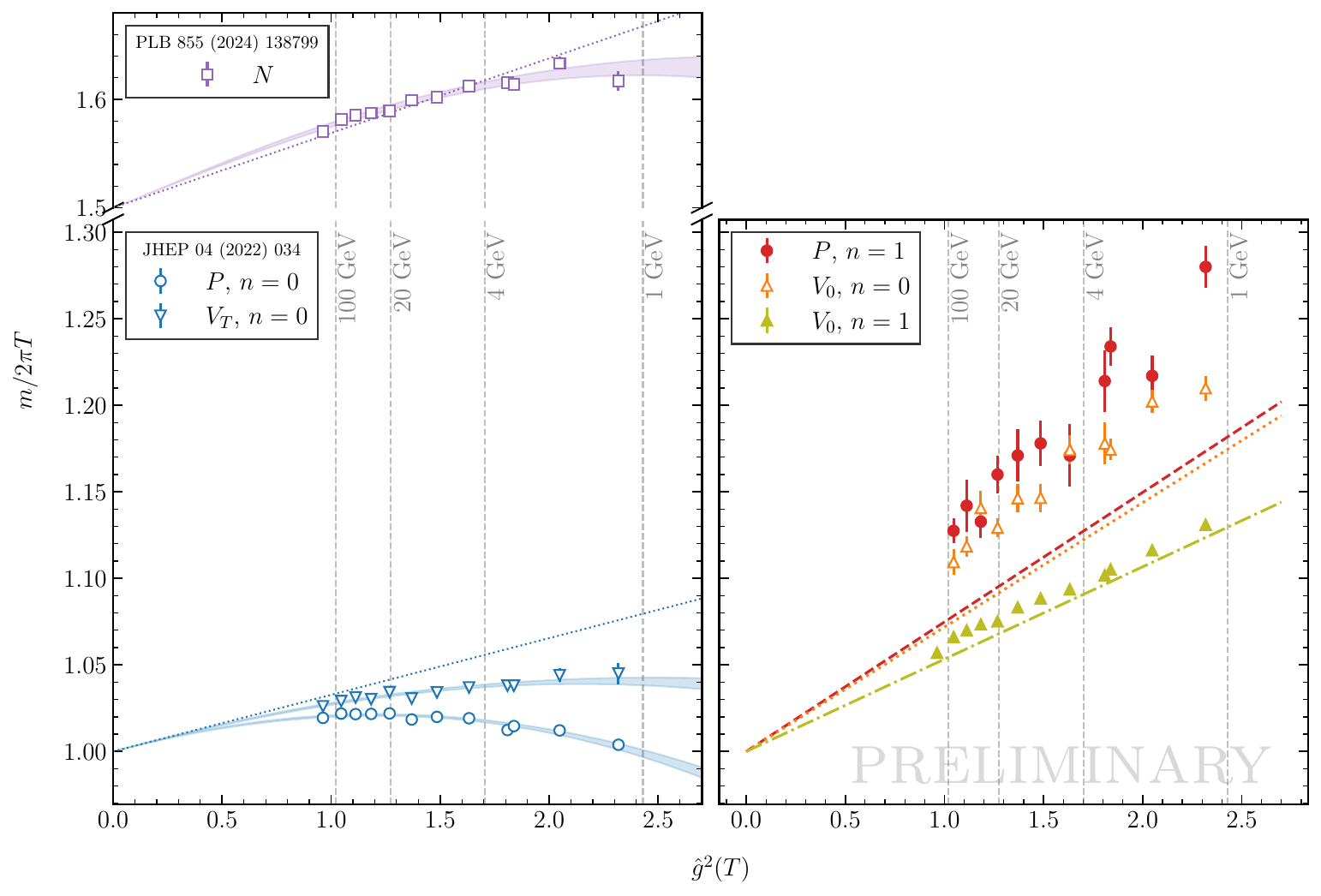}
  \caption{%
    Left: screening masses results as a function of the QCD coupling (and in turn temperature) for pseudoscalar and vector $n=0$ screening masses~\cite{DallaBrida:2021ddx} and baryonic (nucleon) screening masses~\cite{Giusti:2024ohu}.
    Right: preliminary new results presented in these proceedings for additional $n=0$ and $n=1$ mesonic screening masses.
  }\label{fig:summary}
\end{figure}

We have presented the non-perturbative computation of hadronic screening masses in thermal QCD up to $T\approx\SI{160}{\GeV}$.
The plots in figure~\ref{fig:summary} show the newly obtained results discussed in section~\ref{sec:non-static} (on the right) together with previously published results (on the left), including the static Matsubara sector mesonic screening masses~\cite{DallaBrida:2021ddx} and baryonic screening masses~\cite{Giusti:2024ohu}.
We also computed at leading order the hyperfine splitting in the static sector in the dimensionally-reduced effective field theory descriptions of QCD at asymptotically high temperature~\cite{Ce:2025ihd}.
As discussed in section~\ref{sec:hyperfine}, this is compared to the hyperfine splitting measured in non-perturbative simulations of thermal QCD.
While the spectrum of both static and non-static screening masses is qualitatively described by the perturbative expansion originating from the three-dimensional effective theory valid at asymptotically high temperatures, our high-precision numerical computation reveals that even at the highest temperature considered the leading-order perturbative calculation is not compatible with the non-perturbative QCD result.
In particular, the hyperfine splitting measured in QCD is approximately three times larger than the leading $g^4$ contribution.
Terms that are higher order in the QCD coupling are needed, and by including terms proportional to $g^5$ and $g^6$ we are able to describe the hyperfine splitting to per-mille precision for all $T>\SI{1}{\GeV}$.
Coefficients of $\order*{g^5}$ receive contributions from the non-perturbative ultra-soft chromomagnetic sector of high-$T$ QCD and are thus intrinsically not computable with perturbative methods.
The fact that perturbation theory is not enough to fully describe QCD data even at temperatures of the order of the electroweak scale, and that non-perturbative contributions remain sizable, has been observed also in the study of the QCD equation of state~\cite{Bresciani:2025vxw,Bresciani:2025mcu}.

% Future work will focus on $n=2$ sectors.

\section*{Acknowledgements}

The research of M.C.\ is funded through the Italian Ministry of University and Research (MUR) programme for young researchers \enquote{Rita Levi Montalcini}.
This work was (partially) supported by ICSC -- Centro Nazionale di Ricerca in High Performance Computing, Big Data and Quantum Computing, funded by the European Union -- NextGenerationEU.
We acknowledge the EuroHPC Joint Undertaking for awarding this project access to the EuroHPC supercomputer MareNostrum 5, hosted by Barcelona Supercomputing Center (BSC), Spain and the MareNostrum 5 consortium through a EuroHPC Extreme Access call, project no.\ EHPC-EXT-2024E01-033.
We acknowledge the ICSC for awarding this project access to the EuroHPC supercomputer LEONARDO, hosted by CINECA, Italy.

\setlength{\bibsep}{0pt plus 0.3ex}
\bibliography{./bibliography.bib}

\end{document}